# Performance Evaluation Of Direction of Arrival Estimation Using MATLAB

Sai Suhas Balabadrapatruni

Dept. of Electronics & Communication Engineering, JNT University, Hyderabad, India
saisuhas.b@gmail.com

## ABSTRACT

*This paper presents the performance analysis of directions of arrival estimation techniques, Subspace and the Non-Subspace methods. In this paper, exploring the Eigen-analysis category of high resolution and super resolution algorithms, presentation of description, comparison and the performance and resolution analyses of these algorithms are made. Sensitivity to various perturbations and the effect of parameters related to the design of the sensor array itself such as the number of array elements and their spacing are also investigated. The analysis is based on linear array antenna and the calculation of the pseudo spectra function of the estimation algorithms. Algorithms namely Delay-and-sum, Capon's, MUSIC, and ESPRIT Direction Of Arrival Estimates. MATLAB™ is used for simulating the algorithms.*

## 1. INTRODUCTION

The need for Direction-of-Arrival estimation arises in many engineering applications including wireless communications, radar, radio astronomy, sonar, navigation, tracking of various objects, rescue and other emergency assistance devices. In its modern version, DOA estimation is usually studied as part of the more general field of array processing. Much of the work in this field, especially in earlier days, focused on radio direction finding – that is, estimating the direction of electromagnetic waves impinging on one or more antennas [1].

Due to the increasing over usage of the low end of the spectrum, people started to explore the higher frequency band for these applications, where more spectrums is available. With higher frequencies, higher data rate and higher user density, multipath fading and cross-interference become more serious issues, resulting in the degradation of bit error rate (BER). To combat these problems and to achieve higher communication capacity, smart antenna systems with adaptive beam forming capability have proven to be very effective in suppression of the interference and multipath signals [2].

Signal processing aspects of smart antenna systems has concentrated on the development of efficient algorithms for Direction-of-Arrival (DOA) estimation and adaptive beam forming. The recent trends of adaptive beam forming drive the development of digital beam forming systems [3].

Instead of using a single antenna, an array antenna system with innovative signal processing can enhance the resolution of DOA estimation. An array sensor system has multiple sensors distributed in space. This array configuration provides spatial samplings of the received waveform. A sensor array has better performance than the single sensor in signal reception and parameter estimation [4].

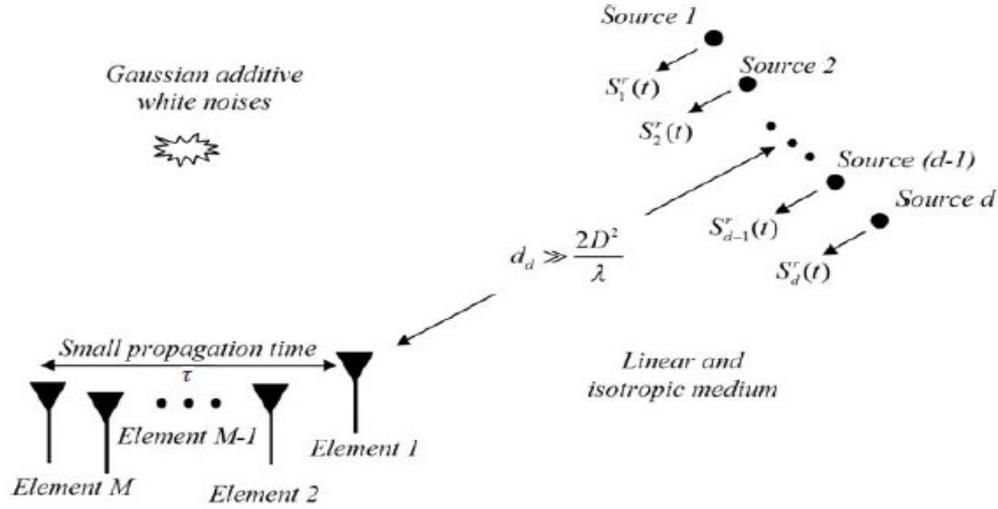

Figure 1. Overview of Direction-Of-Arrival Estimation

## 2. RELATED WORK

Consider an array of 'r' sensors with arbitrary locations and arbitrary directional characteristics, which receive signals generated by 'q' narrowband sources with known centre frequency 'ω' and locations $\emptyset_1, \emptyset_2, \ldots \emptyset_k$. Since the signals are narrowband, the propagation delay across the array is much smaller than the reciprocal of the signal bandwidth and it follows that by using a complex envelope representation, the array output can be expressed as,

$$u(t) = \sum_{k=1}^{q} a(\emptyset_k) s_k(t) + n(t)$$

where,

1. $x(t) = [x_1(t) \ldots \ldots x_r(t)]^T$, is the vector of signals received by the array antenna.
2. $s_k(t)$, is the signal emitted by the $k^{th}$ source as received at the reference sensor 1 of the array.
3. $a(\emptyset_k) = [1, e^{-j\omega\tau_2(\emptyset_k)}, \ldots \ldots e^{-j\omega\tau_r(\emptyset_k)}]^T$, is the steering vector of the array towards the direction $\emptyset_k$.
4. $\tau_i(\emptyset_k)$, is the propagation delay between the first and the $i^{th}$ sensor for a waveform coming from the direction $\emptyset_k$.
5. $n(t) = [n_1(t), \ldots \ldots n_r(t)]^T$, is the noise vector.

## 2.1 Uniform Linear Array Antennas

Consider an M-element uniformly spaced linear array [5] which is illustrated in Figure 2.2. In Figure 2.2, the array elements are equally spaced by a distance d, and a plane wave arrives at the array from a direction $\theta$ off the array broadside. The angle $\emptyset$ is called the direction-of-arrival (DOA) or angle-of-arrival (AOA) of the received signal, and is measured clockwise from the broadside of the array.

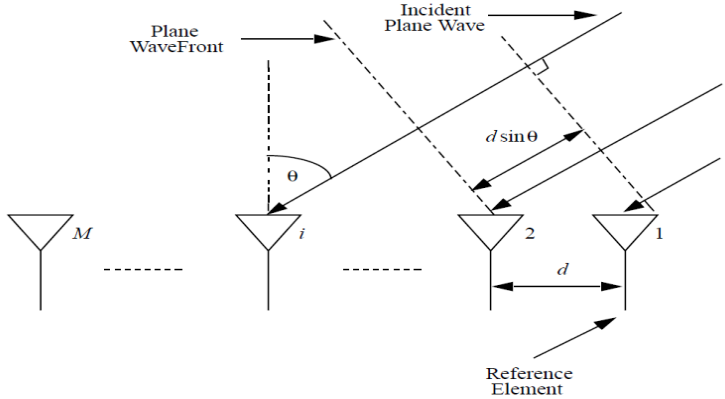

Figure 2. Illustration of a plane wave incident on a Uniformly Spaced Linear Array Antenna from direction $\emptyset$.

Consider the signal be the complex sinusoidal signal which is represented in complex from as, $e^{j\omega_0 t}$. The signal received by the reference element antenna is given by,

$$S_1(t) = e^{j2\pi f_0 t} \tag{1}$$

From the Figure 2, the received signal by the second element of the array antenna will be the delayed version of the signal received by the first element. Consider the delay occurred be 'τ'. Hence, the signal at the second element is given by,

$$S_2(t) = S_1(t - \tau)$$
$$= e^{j2\pi f_0 t}. e^{-j2\pi f_0 \tau} \tag{2}$$

Delay time, 'τ' is given by,

$$\tau = \frac{d \sin \theta}{c}$$
$$= \frac{d \sin \theta}{f_0 \lambda_0} \tag{3}$$

On substituting Eqn. (3) in Eqn. (2), we get,

$$S_2(t) = e^{j2\pi f_0 t}. e^{-j2\pi f_0 \frac{d \sin \theta}{f_0 \lambda_0}}$$

$$= e^{j2\pi f_o t} \cdot e^{-j2\pi \frac{d \sin \theta}{\lambda_o}}$$

$$= e^{j2\pi f_o t} \cdot e^{-j\emptyset} \tag{4}$$

where,

$$\emptyset = 2\pi \frac{d \sin \theta}{\lambda_o} \tag{5}$$

Therefore,

$$S_2(t) = S_1(t) e^{-j\emptyset} \tag{6}$$

The total signals received by the array antenna elements are,

$$X_1(t) = S(t) + n_1(t)$$

$$X_2(t) = S(t - \tau) + n_2(t) \quad \text{And so on.}$$

It is represented in vector form as,

$$\begin{bmatrix} x_1(t) \\ x_2(t) \\ \vdots \\ x_m(t) \end{bmatrix} = S(t) \begin{bmatrix} 1 \\ e^{-j\emptyset} \\ e^{-j2\emptyset} \\ \vdots \\ e^{-j(m-1)\emptyset} \end{bmatrix} + \begin{bmatrix} n_1(t) \\ n_2(t) \\ \vdots \\ n_m(t) \end{bmatrix} \tag{7}$$

⇩

Beam Steering Vector [6]

i.e.,

$$x(t) = \sum_{i=1}^{M} a(\emptyset_i) s_i(t) + n(t) \tag{8}$$

We represent the complex analytical signal as,

$$x(t) = x_I(t) \cos 2\pi f_c t - x_Q(t) \sin 2\pi f_c t \tag{9}$$

Eqn. (9), can be expressed in compact form as,

$$\tilde{x}(t) = x_I(t) + j x_Q(t) \tag{10}$$

Eqn. (10) can be rewritten as,

$$\tilde{x}(t) = r(t) e^{j\emptyset(t)} \tag{11}$$

where,

$$r(t) = \sqrt{x_I^2(t) + x_Q^2(t)} \quad \text{----- Envelope.}$$

$$\emptyset(t) = \tan^{-1}\left(\frac{x_Q(t)}{x_I(t)}\right) \quad \text{----- Phase.}$$

# 3. NON-SUBSPACE TECHNIQUES

These methods depend on spatial spectrum, and DOAs are obtained as locations of peaks in the spectrum. These methods are conceptually simple but offer modest or poor performance in terms of resolution. One of the main advantages of these techniques is that can be used in situations where we lack of information about properties of signal [7].

## 3.1 Delay-And-Sum Method

The delay-and-sum method [8] also referred to as the classical beam former method or Fourier method is one of the simplest techniques for DOA estimation. Figure 2, shows the classical narrowband beam former structure, where the output signal $y(k)$ is given by a linearly weighted sum of the sensor element outputs. That is,

$$y(k) = w^H u(k)$$

(12)

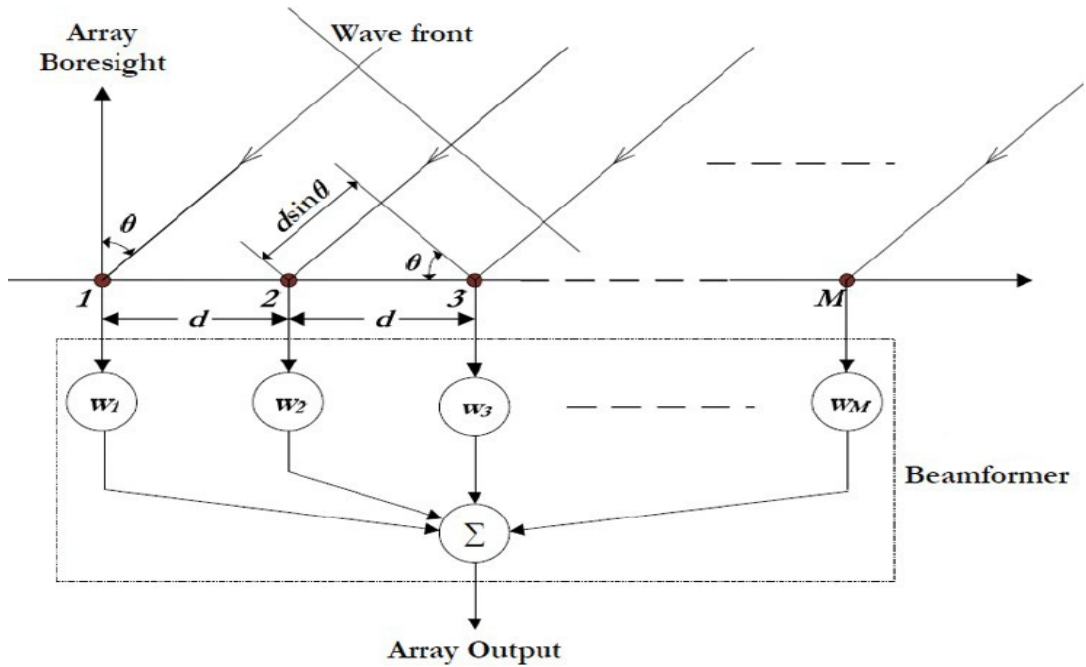

Figure 3. Illustration of Classical Beam Forming

Using equation (12), the output power at the classical beam former as a function of the Angle-Of-Arrival is given by,

$$P_{cbf}(\emptyset) = w^H R_{uu} w = a^H(\emptyset) R_{uu} a(\emptyset) \qquad (13)$$

## 3.2 Maximum Likelihood Technique

Maximum likely hood estimation seeks the parameter values that are most likely to have produced the observed distribution. Maximum likelihood (ML) techniques were some of the first techniques investigated for DOA estimation. Since ML techniques were computationally intensive, they are less popular than other techniques. However, in terms of performance, they are superior to other estimators, especially at low SNR conditions. Moreover, unlike subspace based techniques can also perform well in correlated signal conditions.

Maximum Likelihood (ML) direction-of-arrival (DOA) estimation techniques play an important role in sensor array processing because they provide an excellent trade-off between the asymptotic and threshold DOA estimation performances [9]–[13]. One of the key assumptions used in formulation of both the deterministic and stochastic ML estimators [13] is the so-called spatially homogeneous white noise assumption.

### 3.2.1 Capon's Minimum Variance Technique

The delay-and-sum method works on the premise that pointing the strongest beam in a particular direction yields the best estimate of power arriving in that direction. Capon's minimum variance technique attempts to overcome the poor resolution problems associated with classical beam forming (delay-and-sum method). Using capon's minimum variance method, output power is given by Capon's spatial spectrum,

$$P_{capon}(\emptyset) = \frac{1}{a^H(\emptyset) R_{uu}^{-1} a(\emptyset)} \quad (14)$$

## 4. SUBSPACE TECHNIQUES

Subspace-based methods depend on observations concerning the Eigen decomposition of the covariance matrix into a signal subspace and a noise subspace. Two of these methods MUSIC and ESPRIT were applied here to determine DOA.

### 4.1 MUSIC

MUSIC stands for **Multiple Signal Classification [14].** It is one of the earliest proposed and a very popular method for super-resolution direction finding, which gives the estimation of number of signals arrived, hence their direction of arrival. MUSIC is a technique based on exploiting the Eigen structure of input covariance matrix. Eigen vectors are easily obtained by either an Eigen decomposition of sample covariance matrix or a Singular Value Decomposition (SVD) [15] of the data matrix. In [16], the authors prove that one requires at least K > 2N so that the signal-to-noise ratio is within 3dB of the optimum. While this result cannot be directly applied to the case of DOA estimation, this figure has been taken as a good rule of thumb.

If there are $D$ signals incident on the array, the received input data vector at an $M$-element array can be expressed as a linear combination of the $D$ incident waveforms and noise. That is,

$$u(t) = \sum_{l=0}^{D-1} a(\emptyset_l) s_l(t) + n(t) \tag{15}$$

$$u(t) = [a(\emptyset_0) \quad a(\emptyset_1) \cdots \quad a(\emptyset_{D-1})] \begin{bmatrix} s_0(t) \\ s_1(t) \\ s_{D-1}(t) \end{bmatrix} + n(t) = As(t) + n(t) \tag{16}$$

Where $s^T(t) = [s_0(t), s_1(t) \cdots \cdots, s_{D-1}(t)]$ the vector of incident signals, $n(t) = [n_0(t), n_1(t), \ldots \ldots n_{D-1}(t)]$ is the noise vector, and $a(\emptyset_j)$ is the array steering vector corresponding to the Direction-Of-Arrival of the $j^{th}$ signal. For simplicity, we will drop the time arguments from $u$, $s$, and $n$ from this point onwards.

In geometric terms, the received vector $u$ and the steering vectors $a(\emptyset_j)$ can be visualized as vectors in $M$ dimensional space. From equation (16), it is seen that the received vector $u$ is a particular linear combination of the array steering vectors, with $s_0, s_1, \ldots \ldots s_{D-1}$ being the coefficients of the combination. In terms of the above data model, the input covariance matrix $R_{uu}$ can be expressed as,

$$R_{uu} = E[uu^H] + AE[ss^H]A^H + E[nn^H] \tag{17}$$

$$R_{uu} = AR_{ss}A^H + \sigma_n^2 I \tag{18}$$

where $R_{ss}$ is the signal correlation matrix $E[ss^H]$.

The Eigen values of $R_{uu}$ are the values, $\{\lambda_0, \ldots \ldots \ldots \ldots, \lambda_{M-1}\}$ such that

$$|R_{uu} - \lambda_i I| = 0 \tag{19}$$

Using (18), we can rewrite this as,

$$|AR_{ss}A^H + \sigma_n^2 I - \lambda_i I| = |AR_{ss}A^H - (\lambda_i - \sigma_n^2)I| = 0 \tag{20}$$

Therefore the Eigen values, $v_i$ of $AR_{ss}A^H$ are,

$$v_i = \lambda_i - \sigma_n^2 \tag{21}$$

Since $A$ is comprised of steering vectors which are linearly independent, it has full column rank, and the signal correlation matrix $R_{ss}$ is non-singular as the incident signals are not highly correlated.

A full column rank $A$ matrix and non-singular $R_{ss}$ guarantees that, when the number of incident signals $D$ is less than the number of array elements $M$, the $M \times M$ matrix $AR_{ss}A^H$ is positive semi definite with rank $D$.

From elementary linear algebra, this implies that $M - D$ of the Eigen values, $v_i$, of $AR_{ss}A^H$ are zero. From equation (21), this means that $M - D$ of the Eigen values of $R_{uu}$ are equal to the noise variance, $\sigma_n^2$. We then sort the Eigen values of $R_{uu}$ such that $\lambda_0$ is the largest Eigen value, and $\lambda_{M-1}$ is the smallest Eigen value. Therefore,

$$\lambda_D, \ldots \ldots \ldots, \lambda_{M-1} = \sigma_n^2 \tag{22}$$

In practice, however, when the autocorrelation matrix $R_{uu}$ is estimated from a finite data sample, all the Eigen values corresponding to the noise power will not be identical. Instead they will appear as a closely spaced cluster, with the variance of their spread decreasing as the number of samples used to obtain as estimate of $R_{uu}$ is increased. Once the multiplicity, $K$, of the smallest Eigen value is determined, an estimate of the number of signals, $\widehat{D}$, can be obtained from relation $M = D + k$. Therefore, the estimated number of signals is given by

$$\widehat{D} = M - K \tag{23}$$

The Eigen vector associated with a particular Eigen value, $\lambda_i$ is the vector $q_i$ such that,

$$R_{uu} - \lambda_i I = 0 \tag{24}$$

For Eigen vectors associated with $M - D$ smallest Eigen values, we have

$$(R_{uu} - \sigma_n^2 I)q_i = AR_{ss}A^H q_i + \sigma_n^2 I - \sigma_n^2 I = 0 \tag{25}$$

$$AR_{ss}A^H q_i = 0 \tag{26}$$

Since $A$ has full rank and $R_{ss}$ is non-singular, this implies that

$$A^H q_i = 0 \tag{27}$$

$$\begin{bmatrix} a^H(\emptyset_0)q_i \\ a^H(\emptyset_1)q_i \\ \vdots \\ a^H(\emptyset_{D-1})q_i \end{bmatrix} = \begin{bmatrix} 0 \\ 0 \\ \vdots \\ 0 \end{bmatrix} \tag{28}$$

This means that the Eigen vectors associated with the $M - D$ smallest Eigen values are orthogonal to the $D$ steering vectors that make up $A$.

$$\{a(\emptyset_0), \ldots \ldots \ldots, a(\emptyset_{D-1})\} \perp \{q_D, \ldots \ldots \ldots, q_{M-1}\} \tag{29}$$

This is the essential observation of the MUSIC approach. It means that one can estimate the steering vectors associated with the received signals by finding the steering vectors which are most nearly orthogonal to the Eigen vectors associated with the Eigen values of $R_{uu}$ that are approximately equal to $\sigma_n^2$.

This analysis shows that the Eigen vectors of the covariance matrix $R_{uu}$ belong to either of the two orthogonal subspaces, called the principle Eigen subspace (signal subspace) and the non-principle Eigen subspace (noise subspace). The steering vectors corresponding to the Direction-Of-Arrival lie in the signal subspace and are hence orthogonal to the noise subspace. By searching through all possible array steering vectors to find those which are perpendicular to the space spanned by the non-principle Eigen vectors, the DOA's $\emptyset$'s can be determined.

To search the noise subspace, we form a matrix containing the noise Eigen vectors:

$$V_n = [q_D \quad q_{D+1} \quad \ldots\ldots\ldots \quad q_{M-1}] \qquad (30)$$

Since the steering vectors corresponding to signal components are orthogonal to the noise subspace Eigen vectors, $a^H(\emptyset)V_n V_n^H a(\emptyset) = 0$ for $\emptyset$ corresponding to the DOA of a multipath component. Then the DOAs of the multiple incident signals can be estimated by locating the peaks of a MUSIC spatial spectrum given by,

$$P_{MUSIC}(\emptyset) = \frac{1}{a^H(\emptyset)V_n V_n^H a(\emptyset)} \qquad (31)$$

Or, alternatively,

$$P_{MUSIC}(\emptyset) = \frac{a^H(\emptyset)a(\emptyset)}{a^H(\emptyset)V_n V_n^H a(\emptyset)} \qquad (32)$$

Orthogonality between $a(\emptyset)$ and $V_n$ will minimize the denominator and hence will give rise to peaks in the MUSIC spectrum defined in equation (31) and (32). The $\widehat{D}$ largest peaks in the MUSIC spectrum correspond to the signals impinging on the array.

Once the Direction-Of –Arrival, $\emptyset_i$ are determined from the MUSIC spectrum, the signal covariance matrix $R_{ss}$ can be determined from the following relation,

$$R_{ss} = (A^H A)^{-1} A^H (R_{uu} - \lambda_{min} I) A (A^H A)^{-1} \qquad (33)$$

From equation (33), the powers and cross correlations between the various input signals can be readily obtained.

The DOAs of the multiple incident signals can be estimated by locating the peaks of a MUSIC spatial spectrum given by,

$$P_{MUSIC}(\emptyset) = \frac{1}{a^H(\emptyset)V_n V_n^H a(\emptyset)} \qquad (34)$$

**4.2 Root MUSIC**

Various modifications to the MUSIC algorithm have been proposed to increase its resolution performance and decrease the computational complexity. One such improvement is the Root-MUSIC algorithm developed by Barbell, which is based on polynomial rooting and provides higher resolution, but is applicable only to a uniform spaced linear array. Another improvement proposed by Barbell [17] uses the properties of signal space Eigen vectors (principal Eigen vectors) to define a rational spectrum function with improved resolution capability [Bar83].

**4.3 ESPRIT**

A new approach (ESPRIT) to the signal parameter estimation problem was recently proposed in [18], [19]. ESPRIT is similar to MUSIC in that it exploits the underlying data model and generates estimates that are asymptotically unbiased and efficient. In addition, it has several important advantages over MUSIC.

The ESPRIT method for DOA estimation was first proposed by Roy and Kailath. ESPRIT stands for **Estimation of Signal Parameter via Rotational Invariance Technique [20].** This algorithm is more robust with respect to array imperfections than MUSIC. Computation complexity and storage requirements are lower than MUSIC as it does not involve extensive search throughout all possible steering vectors. But, it explores the rotational invariance property in the signal subspace created by two sub arrays derived from original array with a translation invariance structure. Unlike MUSIC, ESPRIT does not require that array manifold vectors be precisely known, hence the array calibration requirements are not stringent decomposed into two equal-sized identical sub arrays with the corresponding elements of the two sub arrays displaced from each other by a fixed translational (not rotational) distance.

The TLS ESPRIT algorithm is summarized as follows:

1. Obtain an estimate $\widehat{R_{uu}}$ of $R_{uu}$ from the measurements.

2. Perform Eigen decomposition on $\widehat{R_{uu}}$ i.e.,

$$R = V \Lambda V \tag{35}$$

where $\Lambda = diag\{\lambda_0, \lambda_1, \ldots \ldots \lambda_{M-1}\}$ and $V = [q_0, q_1, \ldots \ldots q_{M-1}]$ are the Eigen values and Eigen vectors, respectively.

3. Using the multiplicity $k$, of the smallest Eigen value $\lambda_{min}$, estimate the number of signals $\widehat{D}$, as

$$\widehat{D} = M - K \tag{36}$$

4. Obtain the signal subspace estimate $V = [\widehat{V_0}, \ldots \ldots \widehat{V_{D-1}}]$ and decompose it into sub array matrices,

$$\widehat{V_s} = \begin{bmatrix} \widehat{V_0} \\ \widehat{V_1} \end{bmatrix} \tag{37}$$

5. Compute the Eigen decomposition $(\lambda_1 > \ldots \ldots > \lambda_{2D})$

$$\widehat{V_{01}^H}\widehat{V_{01}} = \begin{bmatrix} \widehat{V_0^H} \\ \widehat{V_1^H} \end{bmatrix} [\widehat{V_0} \quad \widehat{V_1}] = V\Lambda V^H \tag{38}$$

And partition $V$ into $\widehat{D} \times \widehat{D}$ sub matrices,

$$V = \begin{bmatrix} V_{11} & V_{12} \\ V_{21} & V_{22} \end{bmatrix}$$

6. Calculate the Eigen values of $\Psi = -V_{12}V_{22}^{-1}$

where,

$$\widehat{\phi_k} = eigen\ values\ of\ (-V_{12}V_{22}^{-1}) \tag{39}$$

$$\forall k = 0, 1, \ldots\ldots\ldots \widehat{D} - 1$$

7. Estimate the Angle-Of-Arrival as

$$\widehat{\phi_k} = \cos^{-1}\left[ c \frac{\left(arg(\widehat{\phi_k})\right)}{\beta\Delta x} \right] \tag{40}$$

As seen from the above discussion, ESPRIT eliminates the search procedure inherent in most DOA estimation methods. ESPRIT produces the DOA estimates directly in terms of the Eigen values.

## 5. SOURCE ESTIMATION CRITERION

Source detection is of much importance as it determines the location of the signal from where it is arriving in military, these kind of algorithm are employed in identifying the location of the enemy sending the messages. It may be required to find the number of signals being transmitted from some area, so source estimation plays a major role in number of signals incident on the antenna.

I present here two techniques to estimate the number of signals based on the work in [21]. The algorithms start with realizing that the number of signals M is the number of elements

N minus the number of noise Eigen values. These Eigen values are all equal in the ideal case and easy to identify. In practice, due to the estimation of the correlation matrix, the noise Eigen values are not equal, but are close to each other. The algorithms therefore use an estimate of the closeness of the Eigen values.

### 5.1 Akaike Information Criteria

Akaike method [22] does not require any subjective threshold, and the number of sources is determined as the value. In the AIC-based approach, the number of signals $D$, is determined as the value of $d \in \{0,1, \ldots\ldots\ldots\ldots\ldots\ldots\ldots. M-1\}$ which minimizes the following criterion,

$$AIC(d) = -\log \left\{ \frac{\prod_{i=d+1}^{M} \lambda_i^{\frac{1}{(M-d)}}}{\frac{1}{(M-d)} \sum_{i=d+1}^{M} \lambda_i} \right\}^{(M-d)N} + d(2M-d) \tag{41}$$

### 5.2 Minimum Descriptive Length (MDL) Criteria

In the MDL-based approach, the number of signals is determined as the argument which minimizes the following criterion:

$$MDL(d) = -\log \left\{ \frac{\prod_{i=d+1}^{M} \lambda_i^{\frac{1}{(M-d)}}}{\frac{1}{M-d} \sum_{i=d+1}^{M} \lambda_i} \right\}^{(M-d)N} + \frac{1}{2} d(2M-d) \log N_i \tag{42}$$

Here again, the first term is derived directly from the log-likelihood function, and the second term is the penalty factor added by the MDL criterion [23].

| Spatial smoothing Type | $P_{AIC}(d)$ | $P_{MDL}(d)$ |
|---|---|---|
| Forward Only | $d(2M - 2d + 1)$ | $0.5(2M - 2d + 1) \log N$ |
| Forward/Conjugate Backward | $0.5(2M - 2d + 1)$ | $0.25d(2M - d + 1) \log N$ |

Table 1. Modified values of the penalty function in AIC and MDL detection Schemes

These two techniques do not always give the same number of signals. It shows that the MDL approach results in unbiased estimates, while the AIC approach yields biased estimates. In general, therefore, it is better to use the MDL than the AIC approach. Also, recently, Chen et.al, demonstrated an approach based on ranking and selection theory that shows promise [24].

# 6. SIMULATIONS AND RESULTS

## 6.1 MUSIC with Uncorrelated Signal Conditions

The incident signal is obtained from an Omni directional linear antenna array containing 6 antenna elements from which 1024 snapshots are collected. The signals are assumed to be arriving in end fire direction. The elements are separated from each other by distance of about 0.5 in terms of wavelength. The three sources are supposed to be present in the direction $-30°, 0°, 30°$. It is also assumed that noise signals are uncorrelated with each other. The signal to noise ratio is taken as 20 dB. The MUSIC can estimate uncorrelated signals very well. The estimated signal directions are shown in the graph.

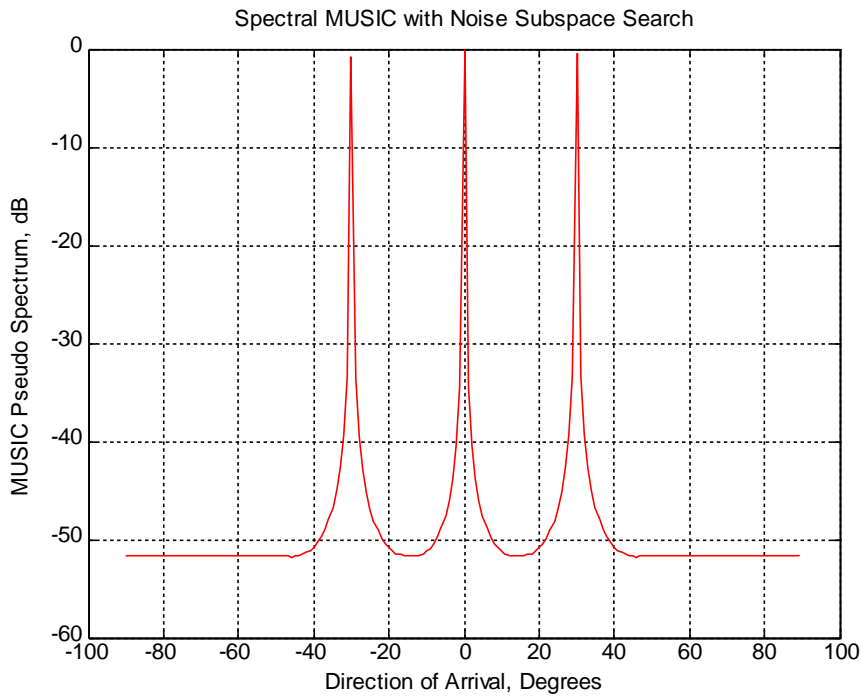

Figure 4. Pseudo Spectrum of MUSIC algorithm with uncorrelated signal conditions

## 6.2 MUSIC with Correlated Signal Conditions

The number of correlated sources present is assumed to be 3 and their positions are assumed to be at angles of $-30°, 0°, 30°$ from the normal direction. The signal to noise ratio is taken as 20 dB. The numbers of input snapshots collected from the 6 element antenna array are 1024. It is assumed that all the elements are separated from each other by a distance of 0.5 in terms of the wavelength. This response is shown in the fig 4,

The MUSIC fails when it comes to detecting correlated input signals. It can be observed from the figure that the response of the MUSIC is not sharp at the peaks while it was sharp in case of uncorrelated input signal condition. This problem arises because the source covariance matrix $R_{uu}$ does not satisfy the full rank condition required by the MUSIC for Eigen decomposition.

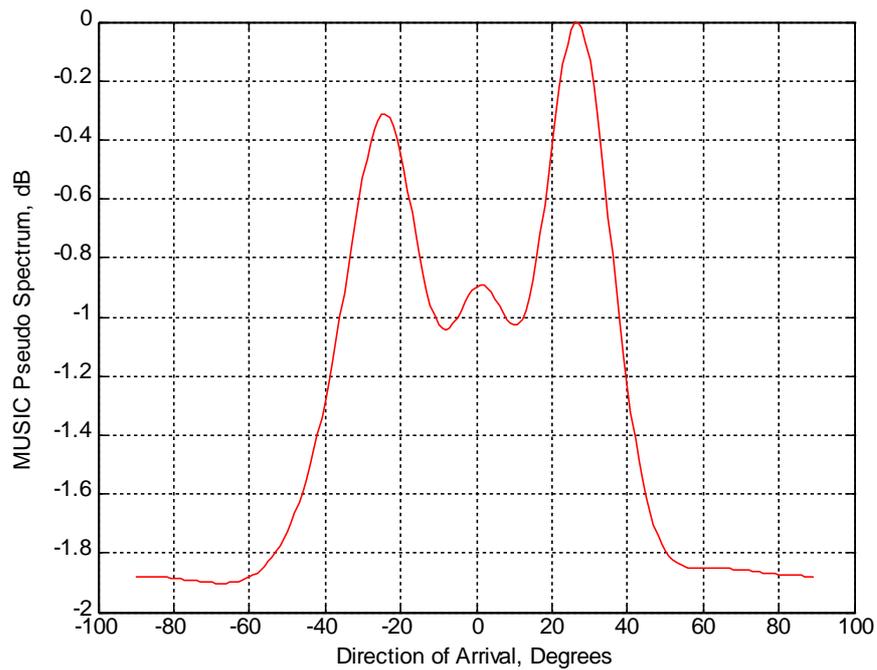

Figure 5. Pseudo Spectrum of MUSIC algorithm with correlated signal conditions

### 6.3 Root MUSIC

The number of array elements used are 10, the inter element spacing in terms of wavelength is given as 0.5. The number of snapshots taken from the antenna is 1024. The directions of the three signals are being radiated are assumed as $-15.5°, -12°, 60.5°$. The noise signals are uncorrelated with each other and the signal to noise ratio is assumed to be 10dB.

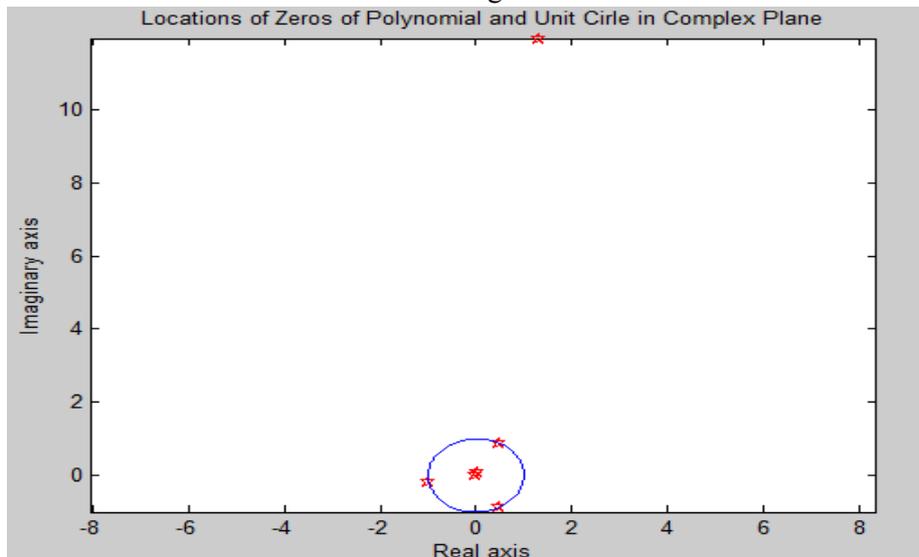

Figure 6. Pseudo Spectrum of Root Music Algorithm

### 6.4 The ESPRIT

The three input signals are assumed to be incident on the antenna array at angle of $-3°, 3°, 61°$. The number snapshots taken from the 6 element antenna array is 1000. The noise signals are all uncorrelated.

The signal to noise ratio is assumed as 12 dB. Here there are two sub arrays each having 3 elements, these elements form a pair of 3 sensor pairs or doublets.

The simulation output is

Number of sources estimated = 3

Direction(s) of arrival estimated in degrees by TLS ESPRIT Algorithm are:

61.00    3.00    -3.02

## 7. Conclusion

In this paper, a novel approach for executing Non-subspace and Subspace DOA algorithms have been studied for different source estimations. Non-subspace estimation techniques yield poor resolution results. When only single source is to be identified, resolution and threshold is not of prime concern. For more than one source, subspace estimation techniques can be used efficiently. Yet the choice of algorithm depends upon the number of sources to be identified and their properties.

The high resolution MUSIC algorithm is based on a single RF port smart antenna has been proposed. After presenting the configuration and the working principle of the antenna, the performance of the proposed technique for various aspects have been studied. The results have justified that the technique for a high-resolution DOA estimation of 1 degree, which is as good as a conventional MUSIC algorithm. However, ESPRIT has proven to be the most accurate method to be used as DOA algorithm. Estimated DOAs were in close agreement to each other.

For radar, DOA estimation is the most important factor to localize targets. For communications, DOA estimation can give spatial diversity to the receiver to enable multi-user scenarios.

Simulation studies have also pointed out that further improvements are needed to enhance system performances to make it more applicable to practical wireless communication systems. The areas of improvements are as follows:

1) The proposed technique works in an uncorrelated signal environment. New methods should explore signal source's spatial signatures in a correlated signal environment.

2) The proposed antenna will not suffer from the negative influence from the mutual coupling between antenna elements. However, the calibration of the antenna aperture over DOA, frequency and temperature, weather environment, and fabrication error is, however, still unavoidable [25] and they influence the antenna performances.

## Acknowledgements

This is an acknowledgement of the intensive drive and technical competence of many individuals who have contributed to the success of my research.

I am grateful to Sri SP DASH, Director, DLRL, Hyderabad and Sri V. Rama Sankaram, Sc ˈ F ˈ chairman, HRDC and members of the HRDC for granting me the permission for the practical training through development of this project in DLRL.

A special note of Thanks to Dr. M. Lakshminarayana, Sc 'G', Gr.Director, COM EW, DLRL who encouraged me in my work.

A special note of Thanks to Sri P.S. Prasad, Sc 'F', Wing Head COM ESM Wing DLRL who encouraged in my work.

I am immensely thankful to Sri K.S.C. Mouleswara Rao, Sc 'F', Division Head COM DF Division for giving me this opportunity and also providing the facilities at ELSEC Division, DLRL.

I am obliged and grateful to my guide Mr. Sanjay Pandav, Sc 'E' of COM DF Division, DLRL for his valuable suggestions, sagacious guidance in all respects during the course of the training.

I would like to express my gratitude to all the members of COM DF Division who were very friendly and cooperative.

I express my profound gratitude and indebtedness to principal, Vignan College for giving me an opportunity to pursue this project.

My sincere thanks are due to Mr. S.S.G.N. Srinivas Rao, Head of ECE Department, also my internal guide, and all faculty members of the college for the encouragement and guidance provided.

**Author**
Sai Suhas Balabadrapatruni

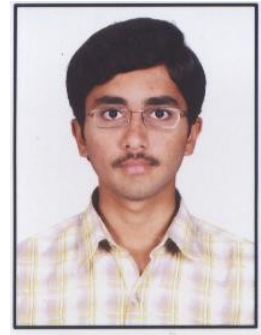

Author Biography

Born in 1990, I am the younger child in my family. Grew up in the south-eastern area of the India with my parents and my elder brother. Completed my schooling in JBMHS and Vignan Vidyalaya. Persuaded my Intermediate (+2) and also completed my Bachelor's Degree from a reputed Organization. Participated in all extra-circular activities like Sports, Cultural, social services etc., in state level as well as district level. All that was 20 years ago! Today I wanted to pursue Master's Degree to continue my research.